\newcommand{\be}{\begin{equation}}
\newcommand{\ee}{\end{equation}}
\newcommand{\bea}{\begin{eqnarray}}
\newcommand{\eea}{\end{eqnarray}}
\newcommand{\ba}[1]{\begin{array}{#1}}
\newcommand{\ea}{\end{array}}

\newcommand{\diracslash}[1]{#1\llap{/\kern2pt}}

\documentclass[12pt]{iopart}
\usepackage{iopams}  
\usepackage{graphicx}
\usepackage{epsfig}
\usepackage{amssymb}
\usepackage{times}
\usepackage{amssymb}
\usepackage{mathrsfs}
\usepackage{graphicx}
\usepackage{epsfig}
\usepackage{dcolumn}
\usepackage{color}
\usepackage{bm}
\begin{document}
\setlength{\topmargin}{0.2in}

\title{Decay dynamics in a strongly driven atom-molecule coupled system}
\author{Arpita Rakshit$^{1}$, Saikat Ghosh$^{2}$ and Bimalendu Deb$^{1,3}$}
\address{$^1$ Department of Materials Science,
$^3$Raman Center for Atomic, Molecular and Optical Sciences,
Indian Association
for the Cultivation of Science,
Jadavpur, Kolkata 700032, India.\\
$^2$ Department of Physics, Indian Institute of Technology, Kanpur, India}

\begin{abstract}
 Within the framework of master equation, we study decay dynamics of an  atom-molecule  system strongly coupled 
by two photoassociation  lasers. 
Summing over the infinite number of electromagnetic vacuum modes that are coupled to the  system, 
we obtain an integro-differential master equation for the  
the system's reduced density matrix. We use this equation to describe correlated spontaneous emission 
from a pair of electronically excited diatomic 
ro-vibrational states. 
The temporal evolution 
of emitted radiation intensity shows quantum beats that result from the laser-induced coherence 
between the two excited states. The phase difference between the two driving fields is found to significantly affect 
the decay dynamics and the beats.  Our results demonstrate the possibility to control decay 
and decoherence in the system  by tuning the relative intensity and the phase between the two lasers. 
We further show that, if the ground-state continuum has a shape resonance at a low energy, 
then the quantum beats  show two distinctive time scales of oscillations in the strong coupling regime. 
One of the time scales originates from the energy gap between the two excited states while
the other time scale corresponds  to the collision energy at which free-bound Franck-Condon overlap 
is resonantly peaked due to the shape resonance. 

\end{abstract}

\maketitle

\section{Introduction} 
Over the last two decades there has been tremendous developments in high precision spectroscopy with cold atoms. It is now possible 
to access low lying rotational levels  of a diatomic molecule formed  by photoassociation (PA) in cold atoms.
For an excited long-ranged molecule 
(formed via narrow-line inter-combination photoassociative transitions as in cold bosonic Sr or  Yb atoms)
the lifetime of excited rotational 
levels can be as large as  10 microseconds. Such metastable molecular excited states are now experimentally accessible using optical spectroscopic techniques. 
This opens up the possibility of creating and studying quantum superposition states between molecular rotational states as well
as superpositions between molecular states and collisional continuum of scattering states between ground-state atoms. 
In a PA process, the scattering state between two ground-state cold atoms become optically coupled to an excited diatomic bound (molecular) state. 
In the weak photoassociative coupling regime, it is interpreted as a loss process and PA spectra are detected in terms of the loss of atoms due to spontaneous  emission from the excited bound state. However, in the strong-coupling regime, the continuum of the scattering states between two atoms becomes strongly-coupled leading to atom-molecule or continuum-bound dressed state quantum dynamics. A transition from the state of two colliding atoms to a diatomic bound state is generally referred to as free-bound transition.  To develop a proper understanding of  correlated quantum dynamics of an atom-molecule coupled system, it is important to formulate a density matrix formalism in 
continuum-bound dressed state picture to appropriately account for spontaneous emission and decoherence in the dynamics.  The influence of 
spontaneous emission on a continuum-bound coupled system had been earlier discussed \cite{agarwal:1982-1984,lewenstein:1983,eberly:1983} 
in the context of autoionizing states. 
Early work  by  Agarwal {\it et al.} \cite{agarwal:1982-1984} treated spontaneous emission of 
continuum-bound coupled  autoionising  Fano state \cite{fano1961} within the master equation 
framework. We adapt such an approach to develop a master equation for the atom-molecule coupled system at ultracold temperatures. Unlike 
most of the the standard systems in quantum optics  dealing  with dissipation and decoherence from discrete levels,
the master equation approach in the present context 
is rather 
involved due to  the continuum of  states of collision between ground-state atoms. 

Here we develop a model to describe coherent effects in an atom-molecule system and 
demonstrate that, after having created coherent superposition between two  rotational states  by strongly 
driving two photoassociative transitions 
with two lasers, the superposition can be detected  as rotational quantum beats in florescence light emitted
from the correlated rotational levels.   Considering $^{174}$Yb as a prototype 
system, we first analyze the ideal situation of the dressed continuum among two exited rotational states and the bare continuum of scattering between ground-state Yb atoms
in the absence of spontaneous emission.  This provides understanding of how the relative intensities and the phases between 
the two driving PA lasers can be used as knobs to manipulate coherence between the two excited states. We then discuss 
the effects of spontaneous emission and decoherence on the dynamical properties of the dressed states.
Our results show that by judiciously 
adjusting relative intensity and the phase between the two lasers it is possible to inhibit spontaneous emission from the two correlated 
excited molecular states and to preserve or manipulate the coherence between the states.  

Quantum beats in radiation intensity arise from coherent superposition of two long-lived  excited states. 
Such state superpositions and their manipulations are of considerable recent interest in quantum information science. 
The possibility of using quantum beats as a spectroscopic measure for quantum superposition was  discussed as early as in 1933 \cite{rmp}.
Experimentally, spectroscopic study of quantum beats started since 1960s \cite{1960s}. The use of lasers to create quantum superposition 
and detect resulting quantum beats in fluorescence started in early 1970s \cite{1970s}.
Forty years ago, 
Haroche, Paisner and Schawlow \cite{schawlow1973}  demonstrated 
quantum beats in florescence light emitted from the excited hyperfine levels of a Cs atom as a signature of 
quantum superposition between the excited atomic states. Since then quantum beats in fluorescence spectroscopy
have been studied in a variety of physical situations \cite{qbeat,qbeat1}. 
These techniques open up new possibilities for studying excited state properties, 
state preparation and manipulation as well as collisional and spectroscopic aspects of ultra-cold atoms and molecules.

\begin{figure}
\includegraphics[width=\columnwidth]{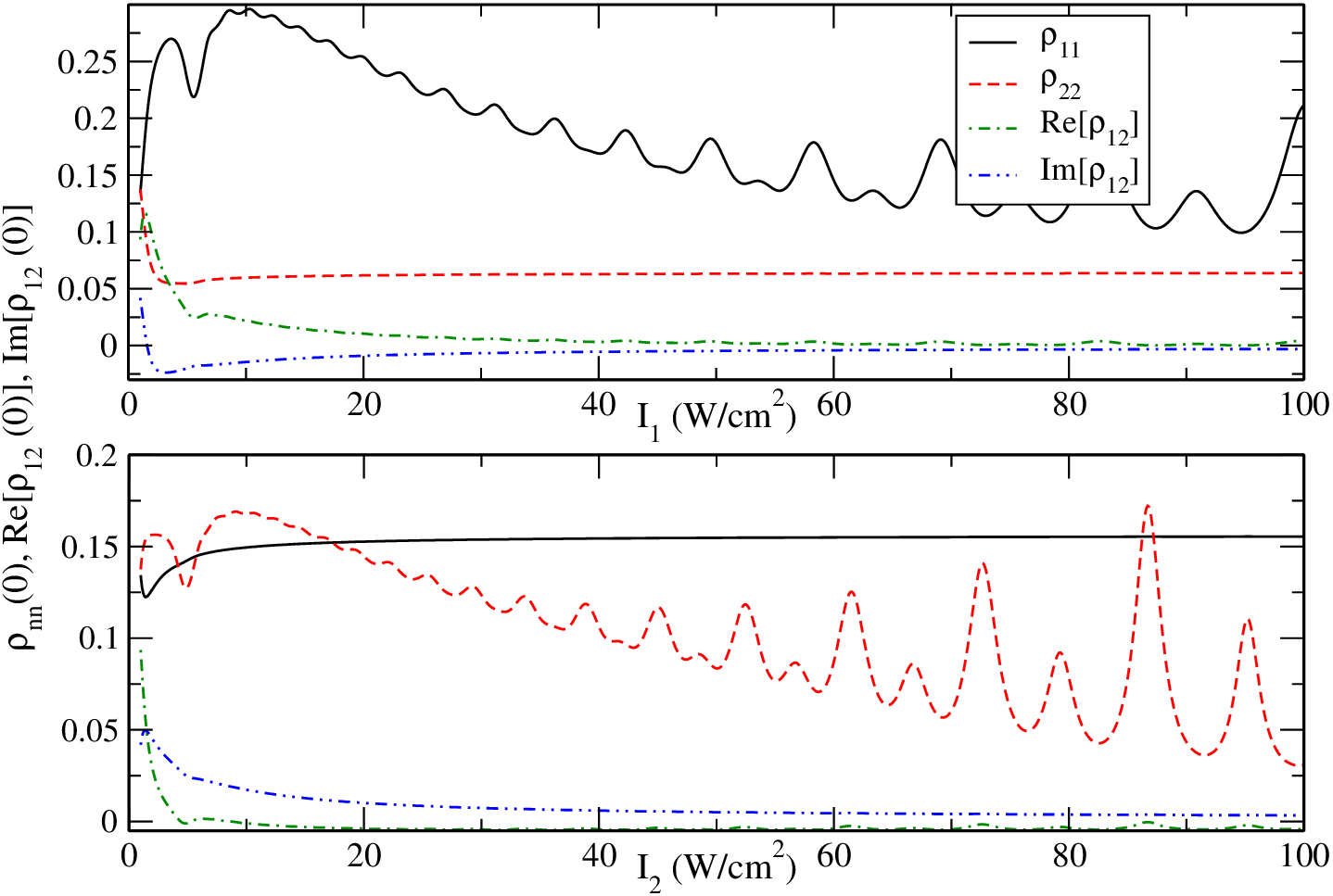}
\caption{(Color online) $\rho_{nn}(0)$ ($n=1,2$)  and the real and  imaginary parts of $\rho_{12}(0)$ are plotted  against $I_1$ (upper panel) 
and $I_2$ (lower panel) in unit of W cm$^{-2}$, keeping the intensity of the other laser fixed 
at 1 W cm$^{-2}$. The other parameters are $\phi=0 $ and $\delta_1 = \delta_2 = 0$.}
\label{rho_i}
\end{figure}

The paper is organized as follows: In section 2, the model
is presented and discussed. We develop a master equation approach 
to spontaneous emission in continuum-bound atom-molecule coupled system in section 3. A solution for the master equation is 
presented. Numerical results 
are analyzed in section 4. The paper is concluded in section 5.

\begin{figure}
\includegraphics[width=\columnwidth]{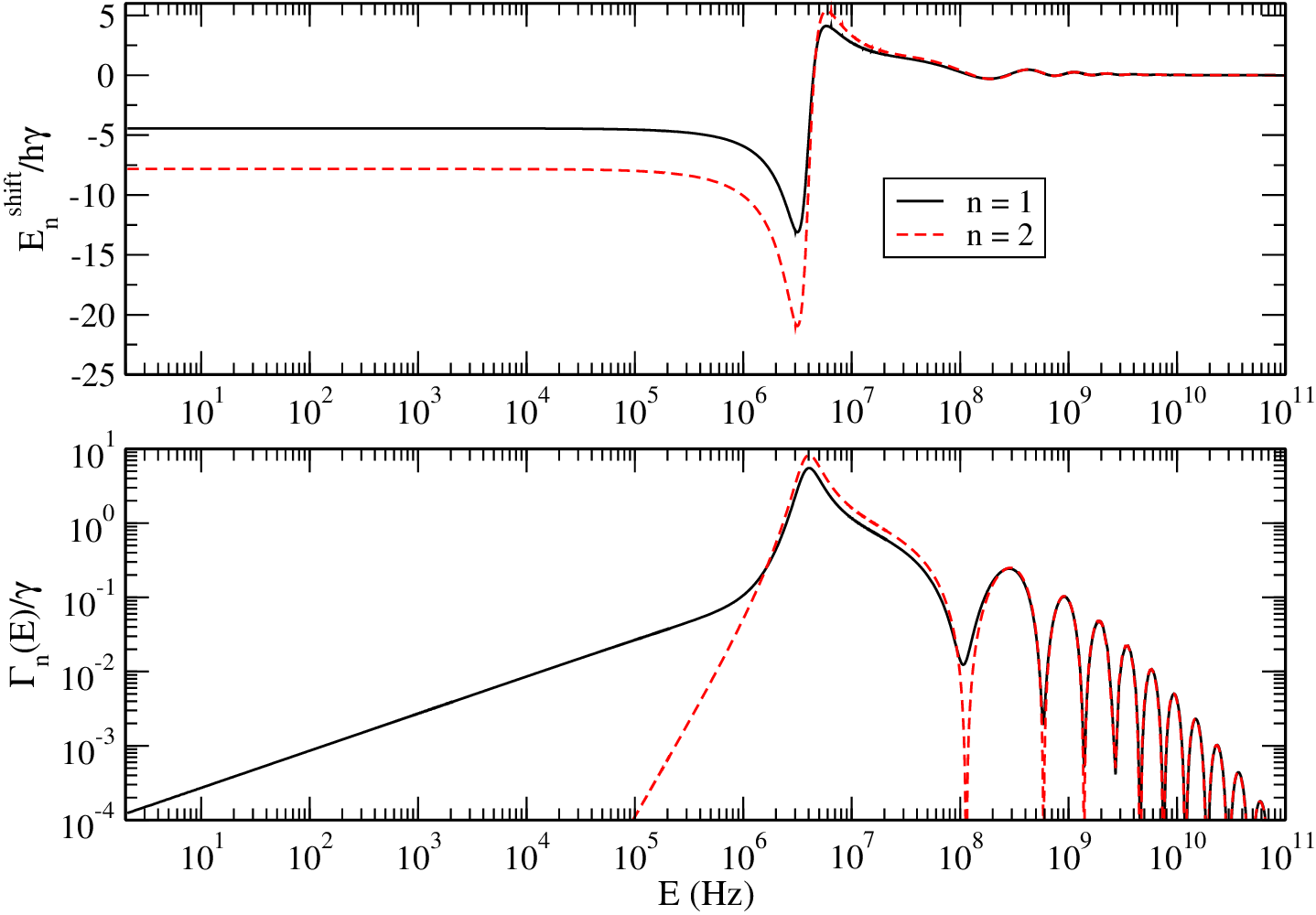}
\caption{ (Color online) Light shifts (scaled by $\hbar \gamma$)  and free-bound stimulated line widths (scaled by $\gamma$)  of the 
two excited bound states $n=1$ ($J=1$) (solid) and $n=2$ ($J=3$) (dashed)  - both having the same vibrational quatum number  $v=106$ of $^{174}$Yb$_2$  (see text)
 are plotted as a function of collision energy $E$ (in Hz)  in  upper and lower panels, respectively; for  
 $I_1 = I_2 = 1$ W cm$^{-2}$ and  the detunings $\delta_1 = \delta_2 = 0$. }
\label{width}
\end{figure}

\section{The model} 

Our model consists of two excited diatomic molecular  ro-vibrational states $|b_1\rangle$ and $| b_2\rangle$
(belonging to the same molecular electronic state)  
coupled to the ground-state bare  continuum $|E\rangle_{\rm{br}} $ of scattering states, by the lasers 1 and 2, respectively. Initially either $|b_1\rangle$ or  $|b_2\rangle$ or 
partially both are populated due to two photoassociation lasers $L_1$ and $L_2$ of
frequencies $\omega_{L1}$ and $\omega_{L2}$, tuned near  $\mid E \rangle_{\rm{br}} \rightarrow \mid b_1\rangle$ and 
$\mid E \rangle_{\rm{br}} \rightarrow \mid b_2\rangle$ transitions, respectively. 
The ground continuum is assumed to have only one internal molecular state with only one threshold and no hyperfine interaction.  
 We assume that 
the two free-bound PA transitions between the ground-state continuum and the two excited ro-vibrational states 
are strongly driven so that  the spontaneous emissions from these two bound states to the continuum   
are negligible as compared to the corresponding stimulated ones. 
However, these two driven bound states can spontaneously decay to other bound states in the ground electronic configuration.
The model we describe in this paper may be contrasted with that in \cite{deb1} where two excited ro-vibrational states populated 
by photoassociation from ground-state continuum are assumed to decay to the same continuum only. 
In the present paper, we primarily discuss the creation of laser-induced coherence and its implications in decay dynamics 
within the framework  of master equation approach while the earlier work \cite{deb1} concerns the creation of 
vacuum-induced  coherence (VIC) with a more simplified model that is solvable by Wigner-Weisskopf method. Compared to the model 
used in \cite{deb1}, the present model is more realistic as it considers decay of the system outside the dressed continuum. Moreover, the present 
work shows exciting possibilities of manipulating excited state coherences using the relative phase between two lasers.

The Hamiltonian governing the dynamics of this system can be written as 
$H = H_{S} + H_{SR}$, where $H_S = H_{\rm{coh}} + \hbar \omega_{b_0}\mid b_0 \rangle
\langle b_0 \mid$  
 is the system Hamiltonian with two parts: the first part $H_{\rm{coh}}$ describes  coherent dynamics with the two strong 
 PA couplings. On the other hand, the second part $\mathcal{H}_{SR}$ is the 
 interaction part of the system with a reservoir of vacuum electromagnetic modes.
Explicitly, one can write 
\begin{eqnarray}
H_{\rm{coh}} &=& \sum_{n=1}^2\hbar(\omega_{b_n} - \omega_{L_n}) \mid b_n\rangle\langle b_n\mid + 
\int{E' \mid E' \rangle_{\rm{br}}\hspace{1mm}_{\rm{br}}\langle E' \mid dE' }\nonumber\\
&+& \int{\sum_{n=1}^2  \left\lbrace \Lambda_{nE'}  \hat{S}_{nE'}^\dagger  + 
{\rm H.C.}\right\rbrace dE'}
\end{eqnarray}
\bea 
H_{SR} &=&\sum_{n=1,2} \sum_{\kappa,\sigma} \hat{a}_{\kappa,\sigma} e^{-i (\omega_\kappa + \omega_{L_n}) t}  
{V}_{n0}(\kappa\sigma) \mid b_n \rangle \langle b_0 \mid  + {\rm H.c.}
\eea 
where $H_{SR}$ is  the Hamiltonian describing the interaction of the system  with the reservoir of 
vacuum modes. Here $\hbar\omega_{b_n}$ are the binding energies of the bound states $|b_{n}\rangle ($n$ = 1,2)$; $|E'\rangle_{\rm{br}}$ is 
the bare continuum state. 
In deriving the above Hamiltonian, we have used rotating wave approximation (RWA) 
\cite{rabi:pr:1937}. In RWA,  one works in a frame rotating 
with the frequency of the sinusoidally oscillating field interacting with a two-level system (TLS) and neglects the counter-rotating 
terms that oscillate with the sum of the field and the system frequencies. It primarily relies on two conditions: (i) the system relaxation time is much larger than 
the time period of oscillation of the field and (ii) Rabi frequency or the system-field coupling is much smaller than the transition frequency of TLS. 
These conditions are in general fulfilled in most cases of a TLS interacting with 
a monochromatic optical field and therefore RWA can be regarded as a cornerstone for studying 
quantum dynamics of TLS. Nevertheless,  RWA may break down in case of intense laser fields or short pulses when 
the Rabi frequency or the coupling becomes comparable with the system frequency. Generally, this may happen when the laser intensity is of the order of $10^{12}$ W cm$^{-2}$ 
or higher. In PA experiments the laser intensity is much lower, typically in the W cm$^{- 2}$ or kW cm$^{-2}$. Strong-coupling 
regime in ultracold PA can be reached with laser intensities higher than 1 kW cm$^{-2}$ 
but much lower than 1 MW cm$^{-2}$. For driven TLS, corrections beyond RWA and in terms of 
Bloch-Siegert shift \cite{bloch:pr:1940} have been discussed by 
Grifoni and Hanggi \cite{hanggi:physrep:1998}. 
The corrections to RWA can be formulated as a systematic 
expansion in terms of the ratio of Rabi frequency to the field frequency \cite{terzidis:epjb:1999}.  
In case of two coupled TLS, there exists 
a parameter regime where leading order term in the expansion vanishes rendering the next higher order term 
to be significant \cite{terzidis:epjb:1999}. However, such situation does not arise in our case and so RWA 
remains valid. 

With electric dipole approximation, 
the laser coupling $\Lambda_{nE'}$ for the absorptive transition from the bare continuum $|E'\rangle_{\rm{br}}$ 
to the  $n$th excited  bound state $\mid b_n\rangle $ is given by 
\bea 
\Lambda_{nE'} &=& e^{i (\mathbf{k}_{L_n}\cdot \mathbf{R} + \phi_{L_n})} \langle b_n \mid \vec{{D}}_n\cdot {\mathbf E}_{L_n} 
\mid E'\rangle_{\rm{br}}
\eea 
where $ \mathbf{k}_{L_n}$, $\mathbf{E}_{L_n}$ and $\phi_{L_n}$  are the wave vector, 
electric field and phase of the $n$th laser, respectively; $\mathbf{R}$ is the center-of-mass position vector of the two 
atoms and  $\vec{{D}}_{n}$  is the free-bound molecular dipole 
moment  associated with the $n$th  bound state. 
The electric dipole approximation here dictates that $k_{L_n} r <\!< 1$, where 
$r$ is the separation between the two atoms.  We have thus used 
$\exp(i \mathbf{k}_{L_n}\cdot \mathbf{r}) \simeq 1$ in writing the above equation.  The operator $\hat{S}_{nE'}^\dagger = |b_n\rangle _{\rm{br}}\langle E' \mid$ is a raising
operator, $\hat{a}_{\kappa,\sigma}$ denotes the annihilation operator of the vacuum field 
$\vec{E}_{vac}$ and ${V}_{n0}(\kappa\sigma ) = -\langle b_n \mid \vec{{D}}_{n0}\cdot
\vec{E}_{vac}(\kappa)\mid b_0 \rangle $ is the dipole coupling with 
$\vec{E}_{vac}(\kappa) = \left(\sqrt{\hbar\omega_{\kappa}/2\epsilon_{0}V}\right)\vec{\varepsilon}_{\sigma}$,
$\omega_\kappa$ being the wave number, $\vec{{D}}_{n0}$ the transition dipole moment between $n$th excited bound state and 
the ground bound state $\mid b_0\rangle$,  $\sigma$ the polarization of the field and $\sqrt{\hbar\omega_{\kappa}/2\epsilon_{0}V}$ the 
amplitude of the vacuum field and $\hbar \omega_{b_0}$ is the binding energy of the bound state $\mid b_0\rangle$. 
The Hamiltonian 
$\mathcal{H}_{\rm{coh} }$ is exactly diagonalizable \cite{deb1, deb2} in the spirit of Fano's theory \cite{fano1961}. 
The eigenstate of $\mathcal{H}_{S}$ is a dressed continuum  expressed as
\begin{eqnarray}
\mid E \rangle_{\rm{dr}} = \sum_{n=1}^{2}  A_{nE}|b_n\rangle  + \int{C_{E^{\prime}}(E) 
\mid E^{\prime}\rangle_{\rm{br}} dE^{\prime}} \label{dressed-cont}
\end{eqnarray}
with the normalization condition $_{\rm{dr}}\langle E^{\prime\prime}|E\rangle_{\rm{dr}} = \delta(E-E^{\prime\prime})$. The coefficients 
$A_{nE}$ and $C_{E^{\prime}}(E)$ are derived in Ref \cite{deb2}.

By using partial-wave 
decomposition of the bare continuum 
$\mid E' \rangle_{\rm{br}} = \sum_{\ell m_{\ell}} \mid E' \ell m_{\ell'} \rangle_{\rm{br}}$,  
 we have  $ \Lambda_{nE'} = \exp[ i ( \mathbf{k}_{L_n}\cdot \mathbf{R} + \phi_{L_n} )] \sum_{\ell m_{\ell}} \Lambda_{J_n
M_{n}}^{\ell m_{\ell}}(E')$ where $J_n$ and $M_n$ are the rotational 
and the magnetic quantum number, respectively, of the $n$th excited bound state in the space-fixed (laboratory) coordinate system.   
Note that $\Lambda_{J_n M_n}^{\ell m_{\ell}}(E)$ represents amplitude for  free-bound transition  from
$(\ell m_{\ell})$ incident partial-wave state to the $n$th bound state. To denote the amplitude for reverse (bound-free) transition, we use 
the symbol $\Lambda^{J_n M_n}_{\ell m_{\ell}}(E)$. Accordingly, we can write 
$A_{nE} = \sum_{\ell' m_{\ell'}} A_{nE}^{\ell' m_{\ell'}} Y_{\ell, m_{\ell'}}(\hat{k})$ and $C_{E^{\prime}}(E) = \sum_{\ell m_{\ell}} 
\sum_{\ell' m_{\ell'}} C_{E^{\prime},\ell m_{\ell}}^{\ell' m_{\ell'}}(E) Y_{\ell' m_{\ell'}}(\hat{k}) $
where $\hat{k}$ represents a unit vector 
along the incident relative momentum between the two atoms. 
Explicitly, 
\bea
 A_{nE}^{\ell' m_{\ell'}} &=&
 \frac{ e^{i \theta_n} \Lambda_{J_n M_n}^{\ell' m_{\ell'}}(E) + \xi_{n'}^{-1} {\cal K}_{nn'}^{\rm{LL}} 
 e^{i\theta_{n'}} \Lambda_{J_{n'}M_{n'}}^{\ell' m_{\ell'}}(E) }{\xi_n - \xi_{n'}^{-1} {\cal K}_{nn'}^{\rm{LL}} 
 {\cal K}_{n'n}^{\rm{LL}} }, \hspace{0.2cm} n' \ne n \label{ane} \\
C_{E^{\prime},\ell m_{\ell}}^{\ell' m_{\ell'}}(E) &=& \delta_{\ell \ell'} \delta_{m_{\ell} m_{\ell'}} 
\delta(E-E^{\prime})+ \sum_{n =1,2} \frac{A_{nE}^{\ell' m_{\ell'}} \Lambda_{\ell m_{\ell}}^{J_n M_n}(E')}{E-E^{\prime}} 
\eea

 where $\theta_n = \mathbf{k}_{L_n}\cdot \mathbf{R} + \phi_{L_n}$, 
\bea  \xi_n(E) = \hbar(\delta_{nE} + i \Gamma_{n}(E)/2) \eea   
\bea 
\hbar \delta_{nE}  =  E +  \hbar \delta_{L_n} - (E_n + E_{n}^{\rm{shift}})
\eea
 with $E_n $ being the binding energy of $n$th excited bound state measured from
 the threshold of the excited state potential, $E_{n}^{\rm{shift}}$ is the light shift of the $n^{\rm{th}}$ bound state  
and 
$\delta_{L_n} = \omega_{L_n} - \omega_A$ with 
$\omega_{L_n}$ is the laser frequency of $n$-th laser and $\omega_A$ the atomic 
transition frequency. 
The 
two lasers interacting with the system results in  an effective coupling  
\bea 
{\cal K}_{nn'}^{\rm{LL}}  = \left  ({\cal V}_{nn'} - i \frac{1}{2} \hbar {\cal G}_{nn'} \right )
\label{knnpll}
\eea 
between the two bound states  
 where

\bea 
{\cal V}_{nn'} &=& \exp[i(\theta_n - \theta_{n'})]
\sum_{\ell m_{\ell}} {\mathcal P} \int dE' \frac{ \Lambda_{\ell m_{\ell}}^{J_n M_n}(E') 
\Lambda^{\ell m_{\ell}}_{J_{n'} M_{n'}}(E') } {E - E'}, \\
{\cal G}_{nn'} &=&  \exp[i(\theta_n - \theta_{n'})]  \frac{2 \pi}{\hbar}  \sum_{\ell m_{\ell}} \Lambda_{\ell m_{\ell}}^{J_n M_n}(E) 
\Lambda^{\ell m_{\ell}}_{J_{n'} M_{n'}}(E).
 \eea 

The term $\Gamma_n(E) = 2\pi|\Lambda_{nE}|^2/\hbar = 2\pi \sum_{\ell, m_{\ell}}
|\Lambda_{\ell m_{\ell}}^{J_n M_n}(E')|^2/\hbar $, is the stimulated linewidth 
of the $n$-th bound state due to continuum-bound laser coupling. Note that the light shift $E_n^{\rm{shift}} = \sum_{\ell} E_{n\ell}^{\rm{shift}}$ is 
the sum over all the partial light shifts  
\bea 
E_{n \ell}^{\rm{shift}} = \sum_{ m_{\ell}} {\mathcal P} \int dE' \frac{ \Lambda_{\ell m_{\ell}}^{J_n M_n}(E') 
\Lambda^{\ell m_{\ell}}_{J_{n} M_{n}}(E') } {E - E'}. \eea 

\begin{figure}
\includegraphics[width=\columnwidth]{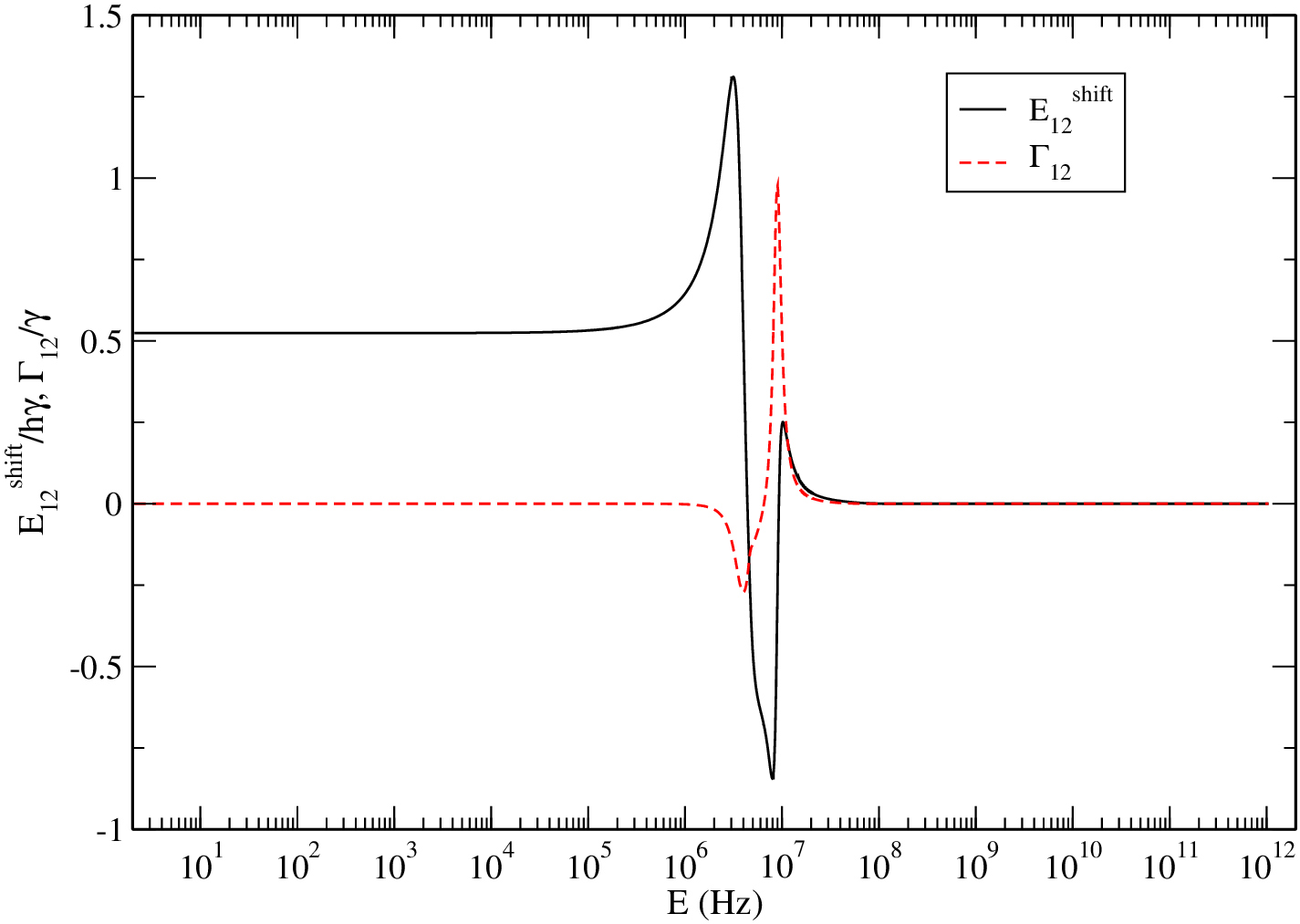}
\caption{(Color online) Plotted are $E_{12}^{\rm{shift}}$  in unit of $h\gamma$ (black solid) and 
$\Gamma_{12}$ in unit of $h\gamma$ (red dashed) are plotted  as a function of collision energy $E$ (in Hz). 
Other parameters are as same as in figure \ref{width}. }
\label{mutual}
\end{figure}

\begin{figure}
\includegraphics[width=\columnwidth]{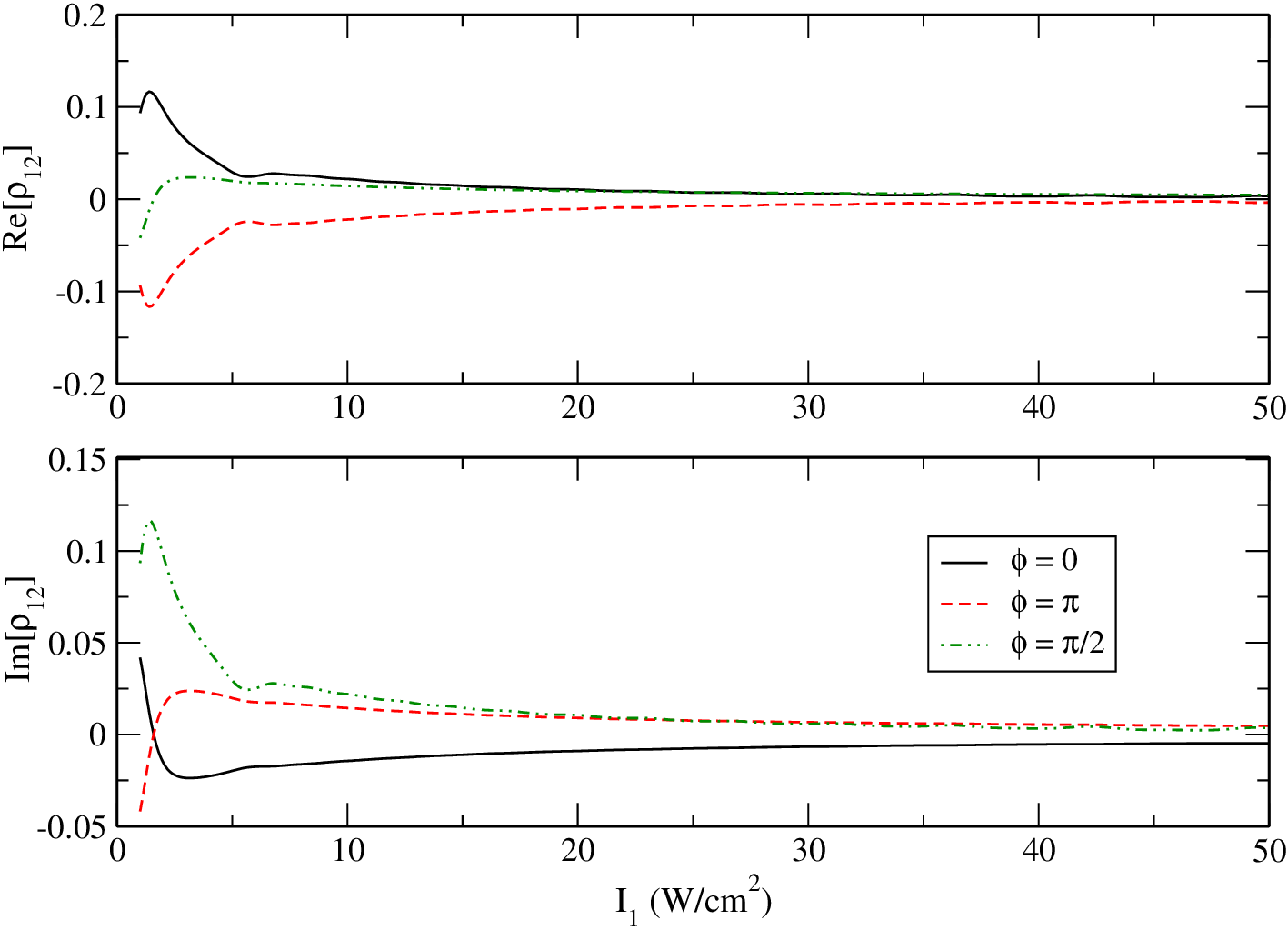}
\caption{(Color online) Re$[\rho_{12}(0)]$ and Im$[\rho_{12}(0)]$ are plotted  against $I_1$ (in unit of W cm$^{-2}$ ) 
for different values $\phi$ of the difference between the phases of the two lasers in upper and lower panels, respectively. The other parameters are $I_2 = 1$ W cm$^{-2}$ and $\delta_1 = \delta_2 = 0$}
\label{real_imaginary_i1}
\end{figure}

\section{Master equation} 

The  system  Hamiltonian can be written in dressed basis as
\bea 
H_0 = \int E d E \mid E \rangle_{\rm{dr}} \hspace{0.1 cm} _{\rm{dr}}\langle E \mid  + \hbar \omega_{b_0} \mid b_0 \rangle \langle b_0 \mid 
\eea 
To derive master equation we work in the dressed continuum basis of the system Hamiltonian.  We  express  bare basis in terms of dressed basis as follows 
\bea 
\mid b_n \rangle =  \int d E \mid E \rangle_{\rm{dr}} \hspace{0.05cm} _{\rm{dr}}\langle E  \mid b_n \rangle =  
\int d E A_{nE}^* \mid E \rangle_{\rm{dr}}
\eea
\bea 
\mid E' \rangle_{\rm{br}} &=&  \int d E \mid E \rangle_{\rm{dr}} \hspace{0.05cm} _{\rm{dr}} \langle E \mid  E'  \rangle\nonumber\\ &=& 
 \int d E C_{E'}^* (E) \mid E \rangle_{\rm{dr}}
\eea 
By substituting all bare basis states with there expansions in terms of dressed basis, we can write system-reservoir interaction Hamiltonian
in terms of dressed basis.  In the interaction picture,  the effective system-reservoir interaction 
Hamiltonian $ H_{SR}^{I} = e^{i H_0 t /\hbar} H_{SR} e^{-i H_0 t /\hbar} $
of the driven system interacting with a reservoir of vacuum 
modes can be written as 
\bea 
H_{SR}^{I} 
&=&  \sum_{\kappa,\sigma} e^{- i \omega_{\kappa} t}   \sum_{n=1}^{2} e^{i(\omega_{b_0} - \omega_{L_n})t} \hat{a}_{\kappa,\sigma}  \int d E 
A_{n E}^*{V}_{n 0}(\kappa \sigma)  e^{i \omega_E t} \hat{S}_{0E}^{\dagger} + {\rm H.c}
\eea

where the superscript `$I$' refers to interaction picture, $ \hat{S}_{0E} = \mid b_0 \rangle \hspace{0.05cm} _{\rm{dr}}\langle E \mid$ and $\omega_{n0} = \omega_{b_n} - \omega_{b_0}$
\begin{figure}
\includegraphics[width=\columnwidth]{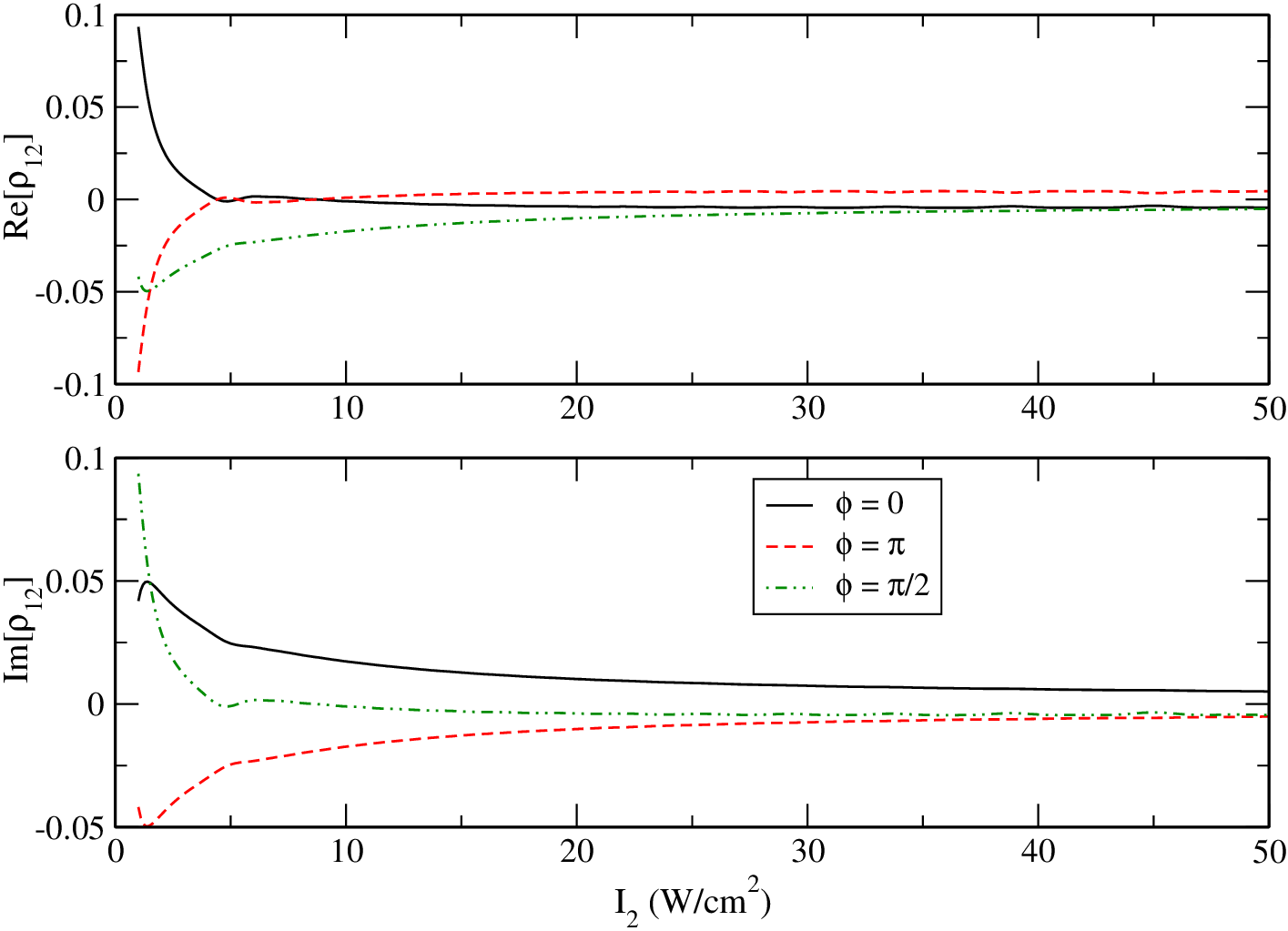}
\caption{(Color online)Same as in figure \ref{real_imaginary_i1}, but as a function of $I_2$ keeping $I_1 = 1$ W cm$^{-2}$ }
\label{real_imaginary_i2}
\end{figure}

Let  ${\rho}_{S + R}(t)$  denote the system-reservoir joint density matrix. 
Following
Agarwal \cite{springer74gsa},  the projection operator ${P}$ is defined by 
\bea 
\mathscr{P} \mathscr{\rho}_{S + R}(t) = \rho_R(0) \rho_{S}(t)
\eea 
wherer     $\rho_{R}$ and $\rho_{S}$ 
are the density matrices of vacuum and the dressed system ($S$) system, respectively.
With the use of this projection operator, Liouville equation under Born approximation can be expressed  \cite{springer74gsa} as 
\bea 
\frac{\partial} {\partial t} \left \{ {P} \rho_{S + R}^{I}(t) \right \} =
- \int_0^t d\tau {P} {L}_{S}^{I}(t) {L}_{S}^{I}(t-\tau)
{P} \rho_{S+R}^{I}(t-\tau) 
\eea
 where 
 \bea 
 \rho^{I} = e^{i H_0 t/\hbar } \rho e^{-i H_0 t/\hbar } 
 \eea 
 is the density matrix in the interaction picture. Here 
\bea 
{L}_{S}^{I}(t) \cdots = \sum_{\kappa,\sigma} e^{-i \omega_{\kappa} t} \left [ \hat{a}_{\kappa\sigma} \hat{\Sigma}_{\kappa\sigma}^{+}(t),\cdots \right ]   
+ \rm{H.c.}
\eea
where 
\bea 
\hat{\Sigma}_{\kappa\sigma}^{+}(t) =  \sum_{n=1}^{2}  e^{-i\omega_{L_n}t}\int d E 
\hat{S}_{0E}^{\dagger}  e^{i(\omega_{E} -\omega_{b_0})t}
A_{n E}^* {V}_{n 0}(\kappa\sigma)
\eea
Tracing over the vacuum states, we obtain
\bea 
\frac{\partial} {\partial t} \left \{  \rho_{S}^{I}(t) \right \} = &-& \sum_{\kappa,\sigma} \int_0^{t}  d\tau \left \{  e^{-i \omega_{\kappa} \tau} 
\left [ \hat{\Sigma}_{\kappa\sigma}^{+}(t),\hat{\Sigma}_{\kappa\sigma}^{-}(t-\tau)\rho_{S}^{I}(t-\tau) \right ] \right. \nonumber\\ &+&
\left.
\left [ \hat{\Sigma}_{\kappa\sigma}^{-}(t),\hat{\Sigma}_{\kappa\sigma}^{+}(t-\tau)\rho_{S}^{I}(t-\tau) \right ] \right \}+ {\rm{H.c.}} 
\label{master}
\eea
From equation (\ref{master}), making use of Markoff approximation, 
we derive the equations of
motion of reduced density matrix elements  in dressed basis. These are 
\bea
\dot{\rho}_{00} &=& \int dE \int dE' \left [\mathscr{A}_{E E'}\rho_{E E'} + \rm{C.c} \right ] \label{rho33a} \\
\dot{\rho}_{E 0} &=&   - i \omega_{E 0}  \rho_{E 0} 
-\int dE' \mathscr{A}_{E' E} \rho_{E' 0} 
- \int dE' \mathscr{A}_{E' E'} \rho_{E 0} 
\label{rhoe3a}\\
\dot{\rho}_{E E'} &=& -i \delta_{E E'} \rho_{E E'}  
-  \int dE'' \mathscr{A}_{E'' E} \rho_{E''E'}  d E' \rho_{E E'}
-  \int dE'' \mathscr{A}_{E' E''}  \rho_{E E''} 
\label{rhosimpa}
\eea
where $ \delta_{E E'} = (E-E')/\hbar$ and 
\bea 
\mathscr{A}_{E E'}  
&\simeq&  \frac{1}{2} \sum_{nn'} \gamma_{nn'}(\omega_n - \omega_{b_0})
\exp[i \delta_{n n'} t] A_{nE} A_{n'E'}^*
\label{aee}
\eea
with $\delta_{n n'} = \omega_{L_n} - \omega_{L_{n'}}$
being the difference between $n$-th and $n'$-th  lasers and 
\bea
\gamma_{nn'}(\omega_n - \omega_{b_0} ) \simeq  \frac{ \vec{D}_{n0} \vec{D}_{0n'} (\omega_n-\omega_{b_0}) ^3}{3 \pi 
\epsilon_0 c^3 \hbar} 
\label{gg}
\eea
$\gamma_{nn`}(x)$ is a function of $x$. 
$\gamma_{nn}$ is the spontaneous linewidth of $n$th excited state and 
$\gamma_{12} =\gamma_{21}$ is the vacuum-induced coupling between the two excited states \cite{Ficek,deb1}.  
Note that in Eq. (27) we have neglected the light shift of the excited levels in comparison to the 
transition frequency  $\omega_{n0} = \omega_n-\omega_{b_0}$ which is in the optical frequency domain while 
the typical light shifts  as shown in Fig.3 are of the order of MHz. The expression (27) 
is obtained in the following way: We first substitute equation (21) into equation (22) and
express the vacuum coupling $V_{n0}$ in terms of corresponding bound-bound transition dipole moment ${\vec D}_{n0}$
as described after equation (3). The sum over $\kappa$ and $\sigma$ is replaced by an 
 integral over the infinite vacuum modes. Using standard Markoffian approximation, one can 
 carry out first the integration over $\tau$ and then over the vacuum modes to arrive at  
 the expression for $\gamma_{nn'}$ as given in equation (27).   The normalization condition is 
\bea 
\rho_{00} + \int \rho_{E E} dE = 1
\label{norm}
\eea 
Equations (\ref{rho33a})-(\ref{rhosimpa}) form a set of three integro-differential equations 
for the  density matrix
elements expressed  in the dressed continuum basis.

\begin{figure}
\includegraphics[width=\columnwidth]{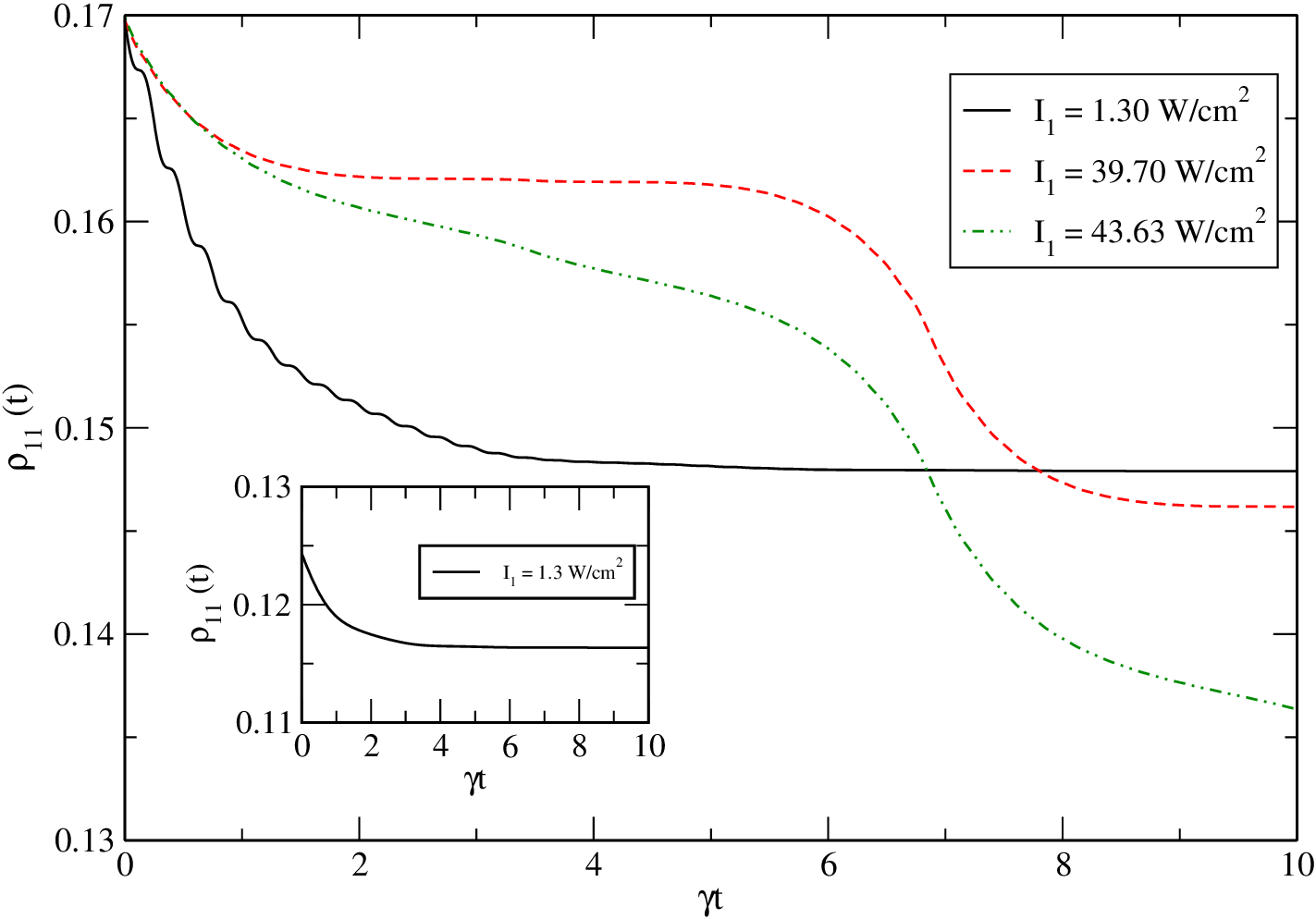}
\caption{(Color online) $\rho_{11}$ as a function of dimensionless unit $\gamma t$ for different values  $ I_1$   
for  $I_2 = 1$ W cm$^{-2}$, $\delta_1 = \delta_2 = 0$ and $\phi =0$. The inset shows the plot of $\rho_{11}$ when laser-2 is switched off and $I_1=1.3$ W cm$^{-2}$, $\delta_1 = 0$ and $\theta_1 =0$. }
\label{rho11-phi=0}
\end{figure}

\section{Solution} 

The density matrix elements can be expressed in bare basis by the transformation
\bea 
\rho_{n n'} = \int d E \int d E' A_{n E} A_{n' E'}^* d E' \rho_{E E'} 
\eea
In interaction picture,  $ {\rho}^I_{E E'} =  \exp(i \delta_{EE'}t) {\rho}_{E E'}$ and  the equation (\ref{rhosimpa}) can be rewritten  as 
\bea
\dot{\rho}^I_{E E'}&=&  -\int dE'' \mathscr{A}_{E'' E} \rho_{E''E'}^Ie^{i\delta_{EE''}t}  
-  \int dE'' \mathscr{A}_{E' E''}  \rho_{E E''}^Ie^{i\delta_{E''E'}t} .
\label{rhofnl}
\eea
The solution of the above equation can be formally expressed as 
\bea
\rho_{E E'}^I(t) &=& \delta(E-E')-\int_{0}^{t} dt'\int {dE'' \mathscr{A}_{E'' E}(t') e^{i\delta_{E E''}t'} \rho_{E''E'}^I (t')}\nonumber\\ 
&-&\int_{0}^{t}
dt'\int {dE'' \mathscr{A}_{E' E''}(t') \rho_{E E''}^I(t') e^{i\delta_{E'' E'}t'}}
\eea
The delta function on the right hand side is the initial value  $\rho_{E E'}^I(0)$. 
The quantity ${A}_{E E'}(t)$ given in equation. 
(\ref{aee}) 
is expressed in terms of the product  $A_{nE}A_{n'E}^*$ of the amplitudes of the $n$th and $n'$th bound states
in energy-normalized 
dressed continuum of equation (\ref{dressed-cont}). If vacuum couplings are neglected, 
the bound-state probability densities are given by $\rho_{nn} = \int 
d E |A_{n E}| d E $ and the coherence terms  $\rho_{nn'} = \int d E  A_{n E} A_{n' E}^*$ with $n'\ne n$. It is important to note that,  
apart from causing spontaneous decay of the $n$th  bound-state probability with decay constant $\gamma_{nn}$, vacuum couplings of the two excited 
bound states $\mid b_1 \rangle$ and $\mid b_2 \rangle$  with the ground bound-state $\mid b_0 \rangle$  effectively give rise to vacuum-induced 
coherence (VIC) \cite{springer74gsa} between the two excited bound states with coupling constant $\gamma_{12}$.   Recently, atom-molecule 
coupled photoassociative systems are shown to be better suited for realizing VIC \cite{deb1}. 
Though the quantities $\gamma_{n n'}$ are calculable from equation (27) when the molecular transition 
dipole moments $D_{n 0}$ are given, for simplicity of our model calculations, we have set 
$\gamma_{11} = \gamma_{22} = \gamma_{12}=\gamma_{21} = \gamma $. In fact, since we consider that both the excited bound states 
 belong to the same vibrational level but differing only in rotational quantum number, the spontaneous linewidths 
 $\gamma_{11}$ and  $\gamma_{22}$ would not differ much. Furthermore, since $\gamma_{12}=\gamma_{21} \simeq \sqrt{\gamma_{11}\gamma_{22}}$, 
 we have $\gamma_{12}=\gamma_{21} = \gamma$ for the case considered here. The stimulated line width $\Gamma_{n}(E)$ 
is a function of the collision energy $E$ for the ground state scattering between the two ground state atoms. Both in the limits
$E \rightarrow 0$ and $E \rightarrow \infty$, $\Gamma_n$ vanishes. Let us fix an energy $\bar{E}$  near 
which both $\Gamma_{1}(\bar{E})$ and $\Gamma_{2}(\bar{E})$ attain their maximum values. It is then possible to write equation 
(\ref{aee}) in the form 
\bea 
\mathscr{A}_{E E'}(t) = \frac{1}{\hbar} \sum_{n n'} \bar{\gamma}_{n n'} \exp[i \delta_{n n'} t]
\bar{A}_{nE} \bar{A}_{n' E'}^*
\eea 
where $\bar{\gamma}_{n n'} =  \gamma/\sqrt{\Gamma_n(\bar{E}) \Gamma_{n'}(\bar{E})}$ and 
$\bar{A}_{nE} = A_{n E} \sqrt{\hbar \Gamma_n(\bar{E})/2} $ are the dimensionless quantities. The absolute value of $\bar{A}_{nE}$ 
is less than unity.
Supposes, the intensities of the two lasers are high enough so that $\Gamma_{n}(\bar{E}) >\!> \gamma$ for both the excited bound states. 
In that case, using $\bar{\gamma}_{n n'}$ or the product $\bar{\gamma}_{n n'} \bar{A}_{nE} \bar{A}_{n' E'}^* $ as a small parameter,
we can  expand equation (\ref{rhofnl})  in a time-ordered series 
\bea
\rho_{E E'}^I(t)&=& \delta(E-E')-\int_{0}^{t} dt'   \mathscr{A}_{E' E} (t')e^{i\delta_{E E'}t'}-\int_{0}^{t} dt'  \mathscr{A}_{E' E} (t')e^{i\delta_{E E'}t'}\nonumber\\&+& \int_{0}^{t} dt'\int dE'' \mathscr{A}_{E'' E} (t')e^{i\delta_{E E''}t'} 
\times \int_{0}^{t'} dt''  \mathscr{A}_{E' E''}(t'') e^{i\delta_{E'' E'}t''}\nonumber\\   
&+& \int_{0}^{t}  dt'\int dE'' \mathscr{A}_{E'' E} (t')e^{i\delta_{E E''}t'}   \int_{0}^{t'}  dt'' \mathscr{A}_{E' E''}(t'') e^{i\delta_{E'' E'}t''} \nonumber \\
&+& \int_{0}^{t} dt'  \int dE'' \mathscr{A}_{E' E''} (t')e^{i\delta_{E'' E'}t'}\int_{0}^{t'} dt'' 
\mathscr{A}_{E'' E}(t'') e^{i\delta_{E E''}t''}   \nonumber\\ 
&+&\int_{0}^{t} dt'  \int dE'' \mathscr{A}_{E' E''} (t')e^{i\delta_{E'' E'}t'} \int_{0}^{t'} dt'' \mathscr{A}_{E'' E}(t'')
e^{i\delta_{E E''}t''} + \cdots  \label{expand}
\eea
It is worthwhile to point out that this method of solution is similar in spirit to that of time-dependent 
perturbation, however it differs in essence because we have used dressed state amplitude  as a small parameter
and not the atom-field coupling. If a large number of terms are taken, then 
the expansion essentially provides solution for any time. However, numerically calculating higher order terms 
becomes increasingly involved because of larger number of multiple integrals in energy variable appearing
in higher order terms. We therefore restrict our numerical studies to a few leading order terms as described 
in the next section. 

\section{Results and discussions}  

\begin{figure}
\includegraphics[width=\columnwidth]{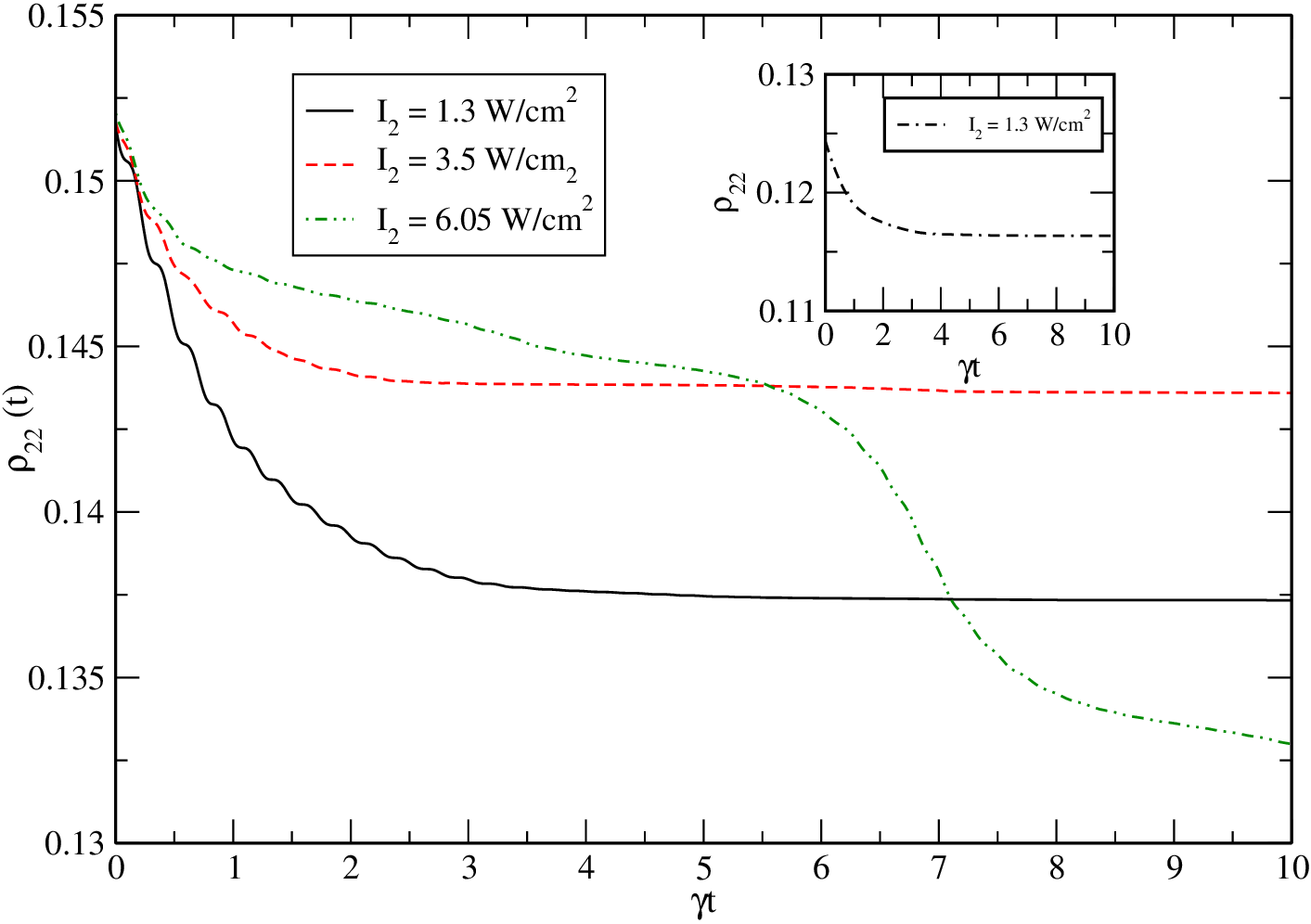}
\caption{(Color online) $\rho_{22}(t)$ is plotted as a function of $\gamma t$  for different values of $I_2$  for $I_1 = 1$  W cm$^{-2}$, $\delta_1 = \delta_2 = 0$ and $\phi =0$.
The inset shows the plot of $\rho_{22}$ when only laser-2 is switched on at intensity $I_2=1.3$ W cm$^{-2}$, $\delta_2 = 0$ and $\theta_2 =0$.}
\label{rho22-phi=0}
\end{figure}

Driven by the two strong lasers, the system  is  prepared in a dressed continuum given by equation (4). Since this state 
is an admixture of the two excited bound states, it is subjected to  
spontaneous emission.  We include spontaneous emission by considering the dressed levels to decay  to 
a third bound level, thereby neglecting the decay of the excited states to the ground-state continuum 
inside the dressed-state manifold.

To discuss the effects of the phase-difference  $\phi = \theta_1 - \theta_2$ between the two lasers, the laser intensities $I_1$
and $I_2$, and the detunings $\delta_1$ and $\delta_2$ on decay dynamics, 
we  first rewrite the dressed-state amplitude $ A_{nE}^{\ell m_{\ell}} $ of equation (\ref{ane})  in the form 
\bea 
A_{nE}^{\ell m_{\ell}} = e^{i \theta_n } \frac{\Lambda_{J_n M_n}^{\ell m_{\ell}}(E) + \mathscr{A}_{n n'}^{\ell m_{\ell}} e^{-i(\theta_n - \theta_{n'})}}{
{\mathscr E}_n + i {\mathscr G}_n/2 } \label{simane}
\eea 
where $\mathscr{A}_{n n'}^{\ell m_{\ell}} =  \xi_{n'}^{-1} {\cal K}_{nn'}^{\rm{LL}} $ and 
\bea 
{\mathscr E}_n = E + \hbar\delta_{n} - ( E_{n} + E_n^{\rm{shift}} + E_{n n'}^{\rm{shift}}), \hspace{0.2 cm}  n' \ne n.  
\eea
The additional shift for the $n$th excited bound state  due to laser-induced cross coupling with the other ($n'$) excited  bound state is 
\bea 
E_{n n'}^{\rm{shift}} = \rm{Re}[  \xi_{n'}^{-1} {\cal K}_{nn'}^{\rm{LL}} {\cal K}_{n'n}^{\rm{LL}} ] 
\eea 
Here  ${\mathscr G}_n = \Gamma_n + \Gamma_{nn'}$ with  
$
\Gamma_{n n'} = - 2 \rm{Im} [  \rm{Re}[  \xi_{n'}^{-1} {\cal K}_{nn'}^{\rm{LL}} {\cal K}_{n'n}^{\rm{LL}} ] 
$ 
being the contribution to the total stimulated line width due to the cross coupling. In expression (\ref{simane}), the first term 
in the numerator corresponds to single-photon transition amplitude due to $n$th laser while the second term describe a net 3-photon 
transition amplitude with 2 photons coming from the $n'$th laser and the other one from $n$th laser.

The foregoing discussion has so far remained quite general.
Now, we apply our method to ultracold $^{174}$Yb atoms. For numerical 
illustration, we  use realistic parameters following the recent experimental 
\cite{tojo:pra:2006,enomoto:prl:2007,enomoto:prl:2008, kitagawa:pra:2008} and theoretical \cite{borkowski:pra:2009} 
works  on PA with $^{174}$Yb. We have chosen  $^{174}$Yb system  
because this offers some advantages compared to other systems. For instance,   
it has no hyperfine structure and the   
ground-state molecular potential of $^{174}$Yb$_2$ is spin-singlet only. Furthermore, it
has spin-forbidden inter-combination transitions. The total rotational quantum number 
is given by $\vec{J}= \vec{J}_e + \vec{\ell}$ where $J_e$ 
 is the total electronic angular momentum.  For numerical work, we specifically consider a pair of 
$^{174}$Yb atoms being acted upon by two co-propagating linearly polarized cw PA lasers. The polarizations 
of both lasers are assumed to be same. This geometry is the same as used in Ref \cite{deb2} for manipulation 
of $d$-wave atom-atom interactions. For our numerical work, we consider that the two lasers  
drive transitions to the molecular
bound states 1 and 2 characterized by  
by the rotational quantum numbers $J_1 = 1$ and $J_2 = 3$, respectively; of the same vibrational level $v=106$.
The two bound states  belong to $0^{+}_{u}$  (Hund's case c) molecular symmetry meaning  
that the projection of $J_e$ on the internuclear axis being zero. Since $^{174}$Yb atoms are bosons, 
only even partial waves are allowed for the scattering
between the two ground state atoms. Free-bound dipole transition selection rules then dictate that the bound state 1 can be accessed from 
$s$- and $d$-wave scattering states, while the bound state 2 is accessible from $d$- and $g$-wave only. Thus $d$-wave ground scattering 
state is coupled to both the excited states by the two PA lasers resulting in  the laser-induced coupling term 
${\mathcal K}_{n n}^{\rm{LL}}$. In general, $d$-wave scattering amplitude is small at low energy. But, fortunately for $^{174}$Yb atoms, 
there is a $d$-wave shape resonance \cite{tojo:pra:2006,enomoto:prl:2007} in the $\mu$K temperature regime leading to  significant enhancement in $d$-wave scattering amplitude 
at  relatively short separations where PA transitions are possible. In our calculations we neglect $g$-wave contributions. 

As we  prepare the system in a desired  dressed continuum, the populations of the two excited bound states and the coherence between them  depend on 
the relative intensity and phase between the two lasers. In the absence of spontaneous emission (idealized situation),
the dressed state properties correspond 
to the initial conditions 
for our model. figure \ref{rho_i} shows variation of the initial populations  and the coherence  as a function of the  
intensity of either laser for  the intensity of the other laser being fixed at 1 W cm$^{-2}$. For all our numerical work, 
we set the spontaneous line width 
$\gamma = 2.29$ MHz \cite{enomoto:prl:2008}.

\begin{figure}
\includegraphics[width=\columnwidth]{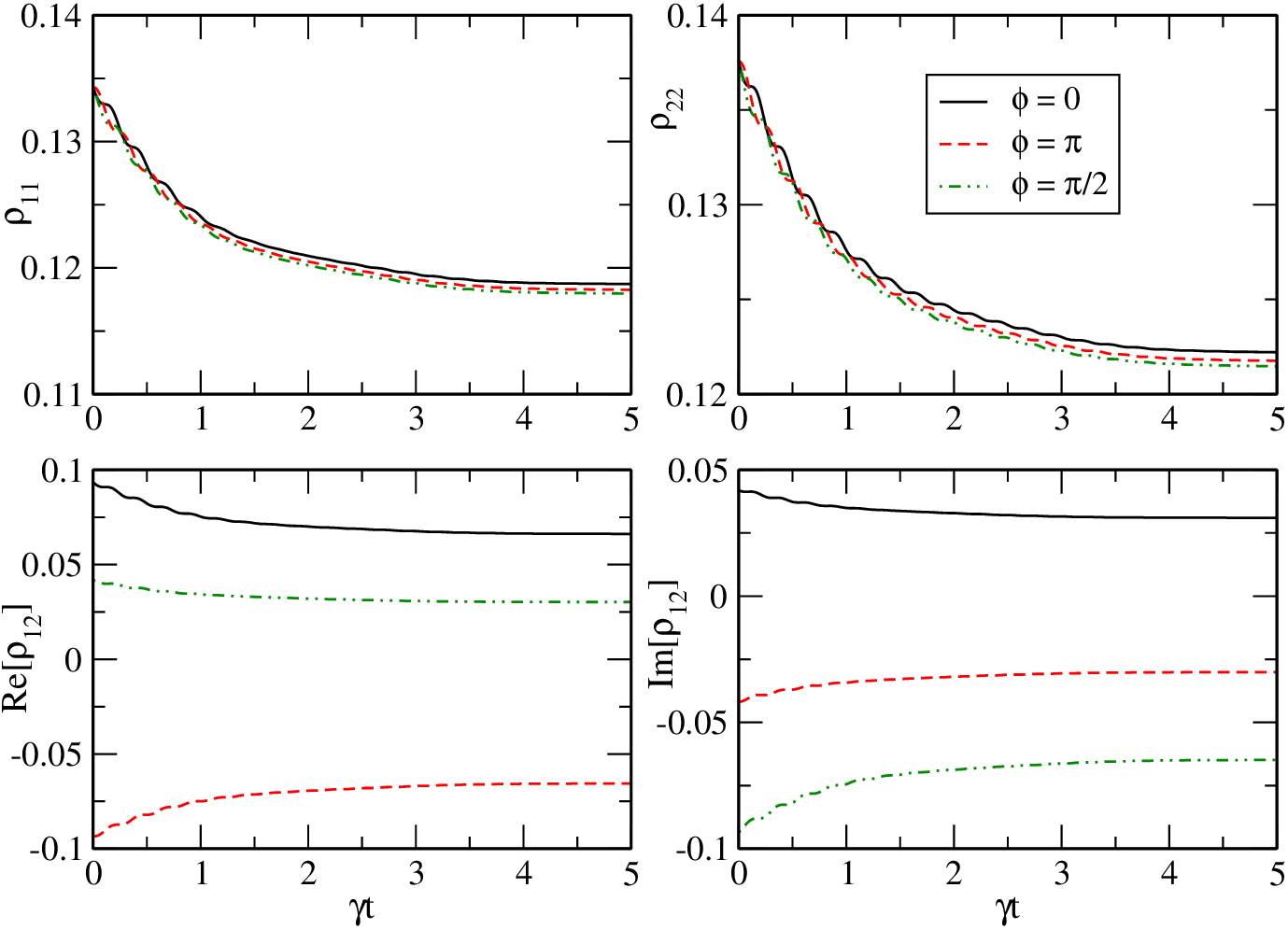}
\caption{(Color online) $\rho_{nn'}(t)$ are plotted against  $\gamma t$ for different values of $\phi$ but for fixed $I_1 = I_2 =1$ W cm$^{-2}$ and $\delta_1 = \delta_2 = 0$.}
\label{rho_nn'-phi}
\end{figure}

The variation of stimulated line widths and light shifts of the two bound states 
of $^{174}$Yb$_{2}$ as a function of collision energy $E$ for the laser intensities $I_1 = I_2 = 1$ W cm$^{-2}$ and 
zero detunings are displayed in figure 2. The shift $E_1^{\rm{shift}}$ (stimulated line width $\Gamma_1$) 
is a sum of $s$- and $d$-wave 
partial shifts ( stimulated line widths) while the shift $E_2^{\rm{shift}}$ and the stimulated line width 
$\Gamma_2$  
are made of mainly $d$-wave partial shift and width, respectively; with no contribution from $s$-wave.   
From figure 2 we notice that the shifts of both bound states  as a function of energy change
rapidly from negative to positive value near  $E = \bar{E} =$ 194 $\mu$K and the stimulated line widths of 
both  bound states exhibit prominent peaks at that energy.  This can be attributed to  a $d$-wave shape resonance 
\cite{tojo:pra:2006,borkowski:pra:2009}.
We have found that 
the $d$-wave partial stimulated line widths of both 
the bound states near shape resonance are comparable. For the first bound state,  
the value of the $d$-wave partial stimulated line width near the resonance is found to   
exceed the $s$-wave partial line width by about 2 orders of magnitude.
 In figure \ref{mutual}, mutual light shift $E_{12}^{\rm{shift}}$ and stimulated line width $\Gamma_{12}$  
 due to the coupling between the 
bound states are shown as a function of collision energy $E$ in Hz. The mutual shifts and widths arise from the coupling 
of the $d$-wave scattering state with the two bound states by the two lasers. Owing to the existence of 
the $d$-wave shape resonance, the laser couplings of the $d$-wave scattering state to both bound states become 
significant, and so are the mutual shifts and stimulated line widths. 

Figures \ref{real_imaginary_i1} and \ref{real_imaginary_i2}  exhibit intensity-dependence of the coherence   $\rho_{12}(0)$. 
The purpose of plotting these two  figures is to assert that it is possible to prepare the
dressed system with a desirable coherence between the two excited bound states 
by judiciously selecting  relative intensities and phases between the two lasers. It is interesting to note that $d$-wave shape resonance 
has a drastic effect on the properties of dressed continuum. Because of this resonance, the $d$-wave contributions 
to the amplitudes of transition to both the bound states are large even at a low temperature allowing an appreciable cross coupling 
to develop between the two bound states.  
For very  large laser intensities at $\delta_{1} = \delta_2 =0 $,
light shifts would be so large that the system will be effectively far off resonant and 
therefore $\rho_{n n'}(0) \simeq 0$. 

For calculating time-dependence of the density matrix elements for all the times, we need to calculate a large number of terms appearing 
on the right hand side 
of equation (\ref{expand}) order by order in $\bar{\gamma}_{nn'}$. This is a laborious and time-consuming exercise. Instead, 
to demonstrate the essential dynamical features  arising from quantum superposition of the two rotational states,
we restrict our study of  decay dynamics to relatively short times. Inserting equation (\ref{aee}) in equation (\ref{expand}), 
retaining the terms up to first order in $\bar{\gamma}_{nn'}$,  we have 
\bea 
\rho_{nn}(t) 
&=&\tilde{A}_{nn}(0) -\gamma\int_{0}^{t} dt' \left[ |\tilde{A}_{12}(t')|^2  + |\tilde{B}_{12}(t')|^2 + |\tilde{A}_{nn}(t')|^2 + 
|\tilde{B}_{nn}(t')|^2 \right ] \nonumber\\ &-&  2 \gamma\int_{0}^{t} dt' \rm{Re} \left \{ \tilde{A}_{12}(t')\tilde{A}_{nn}(t')+ 
\tilde{B}_{nn}(t')\tilde{B}_{12}(t') \right \} \cos(\delta_{12}t')
  + \cdots \label{rhonn}        
\eea
where $\tilde{A}_{nn'}(t)= \int{dE A_{nE}^*A_{n'E}\cos(\omega_Et)}$ and $\tilde{B}_{nn'}(t)= \int{ dE A_{nE}^*A_{n'E}\sin(\omega_E t)}$ 
with $\omega_E = E/\hbar$. Here $\tilde{A}_{nn}(0) = \int dE |A_{nE}|^2 = \rho_{nn}(0)$. 
Similarly, the coherence term $\rho_{12}$ can be calculated up to the first order in $\bar{\gamma}_{n n'}$. These solutions 
 hold good for $\gamma t < |\tilde{A}_{n n}(0)|^{-2}$ or equivalently, $\gamma t < |\rho_{n n}(0)|^{-2}$ for both $n=1,2$. 
 
The decay dynamics of  the populations $\rho_{11}(t)$ and $\rho_{22}(t)$ as a function of the  scaled time $\gamma t$
are shown in figures \ref{rho11-phi=0}  and \ref{rho22-phi=0}, respectively. 
These results  clearly exhibit that, when the system is strongly driven by two lasers, 
the decay is non-exponential and has small oscillations. The oscillation are particularly prominent for short times.  
In  the long time limit  the oscillations slowly die down. However, the oscillations can persist for long times if couplings
are stronger. 
We have chosen the values of the laser intensities $I_1$ and $I_2$ such that the initial values of dressed population  $\rho_{11}(t=0)$ or  $\rho_{22}(t=0)$  are the same 
for those intensities. We notice that, though  the values  $\rho_{11}(0)$   (or $\rho_{22}(0)$)  for a set of $I_1$ values for a fixed $I_2$ value 
(or a set of  $I_2$ values for a fixed $I_1$) are the same, their time evolution is quite different and strongly influenced with the relative intensity 
of the two lasers. That the population oscillations result from the laser-induced coherence between 
the two bound states can be inferred by observing the decay of the populations when either of the lasers is switched off. Plots of $\rho_{11}$ and $\rho_{22}$ against $\gamma t$ for only laser-1 and laser-2  switched on, respectively  are illustrated in the insets of figures \ref{rho11-phi=0}  and \ref{rho22-phi=0} which show exponential decay of the populations $\rho_{11}(t)$ and $\rho_{22}(2)$  with no oscillations. When only one laser is tuned near the resonance 
of either bound state, we do not have any coherence between the two bound states. 
As the two excited bound states are about 57 MHz apart, one of the bound states remains far off-resonant in case of single-laser driving. 
As a result, 
the decay of the driven bound state occurs independent of the other bound state.
The laser-induced coherence between the two bound states is developed only when we apply both the lasers.


The dynamical  characteristics of population 
decay can be interpreted by analyzing the time-dependence and relative contributions of the two expressions  within the third 
and second brackets on the RHS of equation (\ref{rhonn}).  Since $\delta_{12} \simeq - 57 $ MHz and the free-bound couplings are most significant
near $E \simeq \bar{E} \sim 4 $ MHz as can be noticed from figure \ref{width}, we may perform the time integration on the terms associated with 
$\cos(\delta_{12} t)$ in equation (\ref{rhonn}) in the slowly varying envelope approximation to obtain 
\bea
-  2 \gamma \rm{Re} \left \{ \tilde{A}_{12}(t)\tilde{A}_{nn}(t)+ 
\tilde{B}_{nn}(t)\tilde{B}_{12}(t) \right \} \sin(\delta_{12}t)/\delta_{12}
\eea 
Further, since in energy integrations the major contributions will come from energies near $E \simeq \bar{E}$, we may approximate
\bea 
\tilde{A}_{nn'}(t) = \int dE A_{nE}^*A_{n'E}\cos(\omega_E t) \simeq \cos(\omega_{\bar{E}} t) \rho_{nn'}(0)
\eea 
Similarly, $\tilde{B}_{nn'} \simeq \sin(\omega_{\bar{E}} t) \rho_{nn'}(0)$. Using these approximations, we get 
\bea 
\rho_{nn}(t) 
&\sim&\rho_{nn}(0) -\gamma t \left[ |\rho_{12}(0)|^2  + \rho_{nn}(0)^2  \right ] \nonumber \\
 &-&  2 \gamma  \rm{Re} \left \{ \rho_{12}(0)\rho_{nn}(0) \right \} \sin(\delta_{12}t)/\delta_{12}
  + \cdots \label{rhonnapx}        
\eea
This expression clearly shows that when the quantities $(|\rho_{12}(0)|^2  + \rho_{nn}(0)^2)$ and $2 \rm{Re} ( \rho_{12}(0)\rho_{nn}(0) )$ are 
of comparable magnitude, we expect oscillations in population decay with time period $ \tau_{\rm{osc}} = \sim 2\pi/|\delta_{12}| \simeq 0.11$ 
in unit of $\gamma^{-1}$. When the laser intensities are not too high to induce large shifts, we would 
expect the qualitative features of the oscillations will be largely governed by  one time scale which is $\tau_{\rm{osc}}$.
In fact, the solid 
black curves in figures \ref{rho11-phi=0} and \ref{rho22-phi=0} clearly demonstrate oscillatory modulations with 
time scale $\tau_{\rm{osc}}$. 
However, when the laser 
intensities are high enough so that the  energy-dependent shifts and stimulated line widths are appreciable for a range of energies around $E = \bar{E}$, then expression 
(\ref{rhonnapx}) would not be useful to indicate correct qualitative features. In that case, we need to retain full time dependence which will introduce 
another time scale $ 2 \pi/\omega_{\bar{E}}$ which is, in the present context, roughly equal to $2 \pi$ in unit of $\gamma^{-1}$. 
In such situations the net result would be a competition between oscillations with the two time scales. The plots   in figures \ref{rho11-phi=0}
and \ref{rho22-phi=0} at larger 
laser intensity or intensities clearly demonstrate such oscillatory modulations of the population decay with two time scales. It is particularly 
important to note that for larger intensity and appropriate detunings,  the early population decay can be made much slower for an appreciable time duration. It 
is worthwhile to point out that, this analysis is done only to gain insight into the physics of the decay dynamics of the system, all the 
results presented here are obtained by numerically integrating over time $t'$ and the entire range of energy. 

\begin{figure}
\includegraphics[width=\columnwidth]{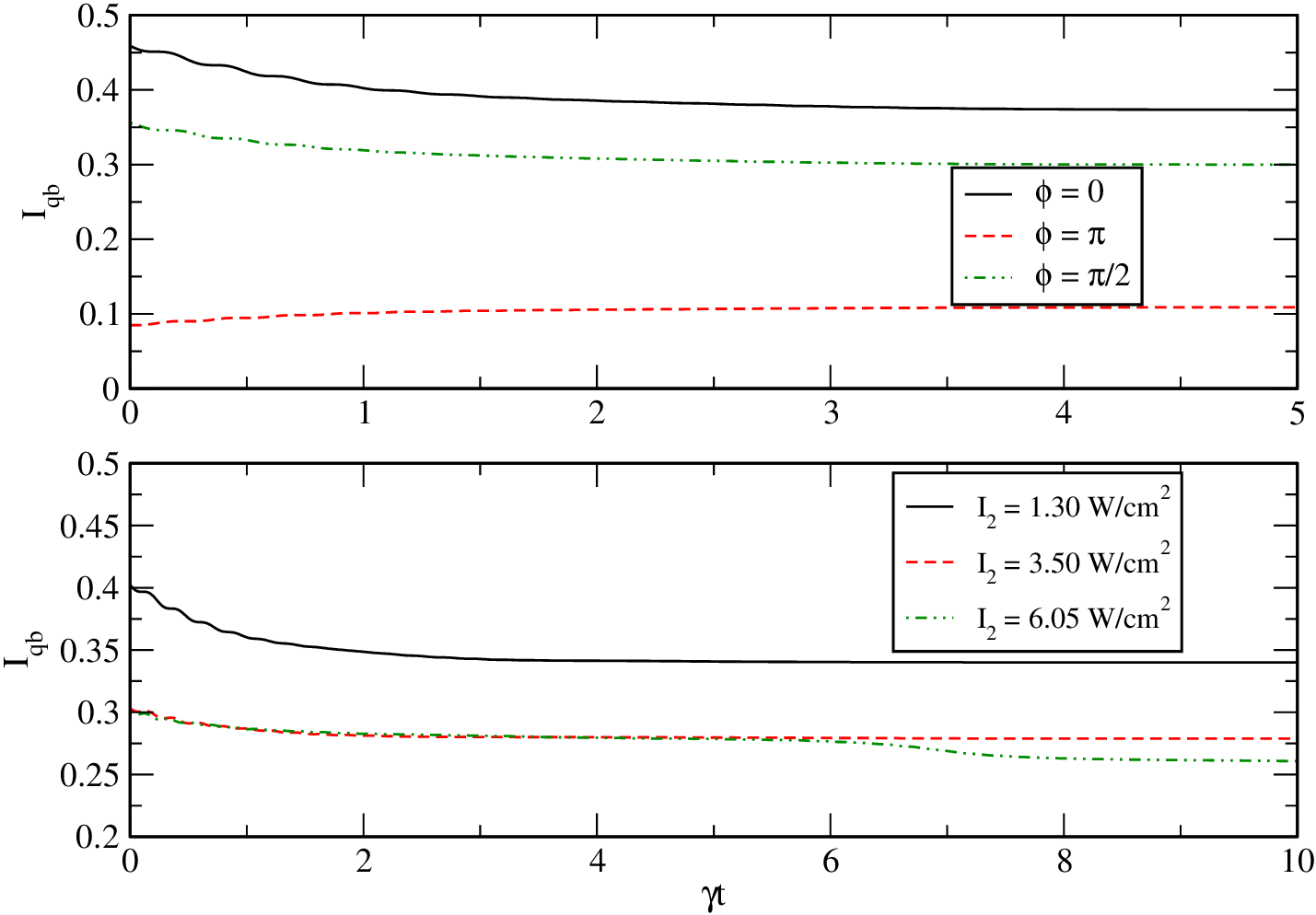}
\caption{(Color online)In upper panel,  we plot $I_{qb}$ against $\gamma t$ for different values of $\phi$ when other parameters are fixed as  $I_1 = I_2 =$ 1 W/cm$^2$ and $\delta_1 = \delta_2 = 0$ MHz. In lower panel  $I_{qb}$ is plotted as a function of $\gamma t$ for different values of $I_2$ for $I_1 =$ 1 W/cm$^2$, $\phi = 0$ and $\delta_1 = \delta_2 = 0$ MHz.}
\label{qb-ph-int}
\end{figure}

Figure \ref{rho_nn'-phi}  shows the effects of  $\phi  $  on 
the temporal evolution of the populations  $\rho_{nn}$  and the 
coherence terms $\rho_{nn'}$ with $n\ne n'$. 
Though  $\phi$ does affect the behavior of the oscillations in  population decay, the magnitude 
of the populations  at a time $t$ is not altered  much with the change of $\phi$. 
In contrast, the magnitudes of the 
real and imaginary parts of the coherence 
term $\rho_{12}$  are largely influenced by $\phi$. When $\phi$ is altered by $\pi$, 
the sign of both real and imaginary parts of $\rho_{12}$ changes. 

\begin{figure}
\includegraphics[width=\columnwidth]{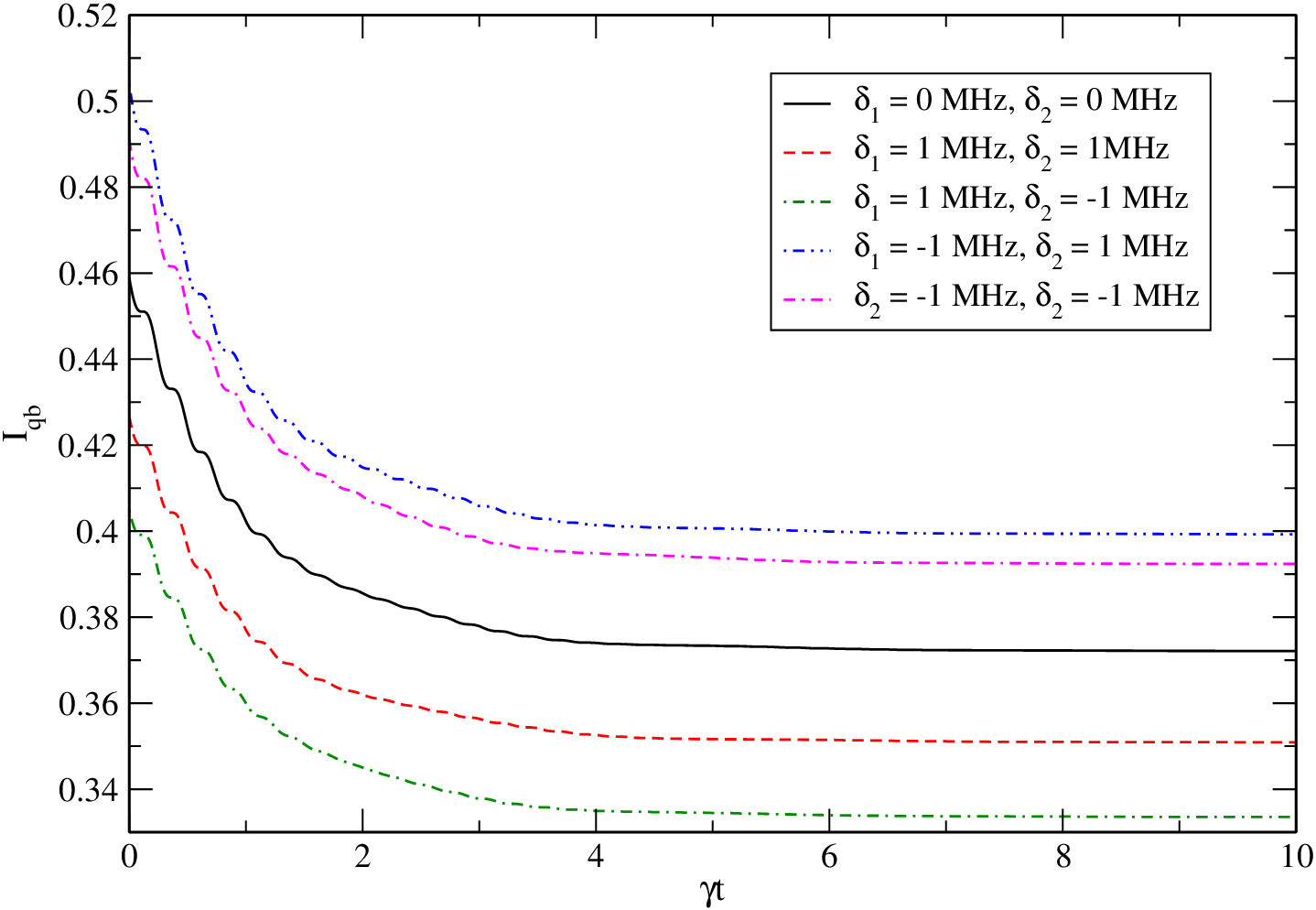}
\caption{(Color online)$I_{qb}$ is plotted against $\gamma t$ for different values of detuning parameters. Other parameters are fixed as  $I_1 = I_2 =$ 1 W/cm$^2$ and $\phi = 0$.}
\label{qb-delta}
\end{figure}

Finally, we discuss quantum beats by studying the temporal evolution of the intensity of light emitted from the two correlated excited bound states.
Quantum beats are manifested as oscillations in the emitted  radiation intensity  $I_{qb}$ as a function of time  
\cite{Ficek:pg 121, Ficek}, which is given by  
\begin{eqnarray}
I_{qb}(t) = \gamma (\rho_{11}(t)+\rho_{22}(t)+ 2\rm{Re}[\rho_{12}(t)])
\end{eqnarray}
In lower panel of figure \ref{qb-ph-int} we show the effects of laser phase $\phi$  on quantum beats in 
time-dependent fluorescent intensity.
The effects of different intensities of laser-2 on quantum beats are illustrated in the lower panel of same figure. 
We demonstrate the effects of different detunings on quantum beats in Fig. \ref{qb-delta}. 
Because of the mutual light shift $E_{12}^{\rm{shift}}$ between the two bound states due to coupling term  
${\mathcal{K}}^{\rm{LL}}_{12}$, the resonance conditions in case of two PA lasers 
are altered in comparison to those in single PA laser case. This  leads to the non-monotonous effects of detunings on quantum beats 
as shown in figure \ref{qb-delta}. 

Before ending this section, we wish to make a few remarks on the possibility of experimental demonstration of 
the physical effects discussed here. Our model can be easily realisable with currently available technology of 
high precision PA spectroscopy. Ultracold bosonic Yb or Sr atoms appear to  be most suitable for this purpose.  Because, they offer several 
advantages. First, their electronic ground state is purely singlet and has no hyperfine structure. This means that the bare continuum has no multiplet 
structure and so there is only one ground-state channel. Second, they have narrow line singlet-triplet inter-combination transitions. Third, they have long-range 
excited bound states which are accessible via PA \cite{borkowski:pra:2009}. These bound states have relatively long life time ($\sim$ microsecond). 
It is possible to selectively drive two rotational levels as required for the model. Furthermore, since both 
the excited bound states have the same vibrational quantum number, their outer turning points will lie  almost at the same separation. Because of long-range 
nature of these excited bound states that are  strongly driven by the two lasers from the bare continuum, 
these two  excited bound states are expected to have the largest Franck-Condon overlap with 
the near-zero energy or the last bound state in the ground electronic potential. It is therefore quite natural that these two driven 
bound states will predominantly spontaneously decay to the last bound state.  In fact, 
the scattering length of Yb atoms have been experimentally determined by detecting the last bound state via two-colour PA spectroscopy \cite{kitagawa:pra:2008}, since the energy 
of the last bound state and the scattering length are closely related. All these facts indicate that our model is a realistic one and the predicted 
excited state coherence and the resulting quantum beats can be experimentally realisable.

\section{Conclusions} 

In conclusion we have developed  a theoretical treatment for the  decay and decoherence from a pair of correlated excited molecular states 
in a strongly driven atom-molecule coupled system. We have shown that it is possible to  create  coherence  
between two excited molecular states by  strong-coupling photoassociation  with two lasers. The transitions between these excited 
states may be electric dipole-forbidden. Our theoretical results have demonstrated that the spontaneous emissions
from these two correlated excited bound states are strongly influenced by the  coherence between them. In 
particular, the spontaneous emission has been shown to occur non-exponentially with multiple time scales due to the coherence. The  
linewidths and light shifts are shown to be largely affected by the coherence leading to the suppression 
of spontaneous emission and decoherence. Our results have  revealed that the coherence can be detected 
as oscillations in decay and decoherence dynamics. We have further shown that the phase-difference between 
the two lasers can be used as a knob to change the phase of these oscillations.
We have also demonstrated quantum beats in fluorescence light 
as a signature of the coherent superposition between the two excited states. We have 
discussed the possibility of experimental realization of our model. 

Finally, our work may be  useful in stimulating further studies on laser manipulation of the continuum and the bound states between 
ultracold atoms. Laser dressing of continuum-bound coupled systems in the context of autoionisation and photoionisation 
 had been extensively studied earlier
demonstrating many interesting  effects such as ``confluence of coherences'' \cite{eberly}, 
population trapping with laser-induced continuum structure \cite{population}, 
non-decaying dressed-states in continuum \cite{slhan}, line-narrowing of 
autoionizing states \cite{linenarrowing}, etc..  Our model can be extended to explore analogous effects in a new parameter regime at the photoassociatve interface \cite{debpramana} 
of ultracold atoms and molecules.

\noindent 
{\bf Acknowledgment} \\
AR is grateful to CSIR, Govt. of India, for a support.

\end{document}